\documentclass[final,5p,times]{elsarticle}

\usepackage{amssymb}
\usepackage{graphicx}
\usepackage{subfig}

\usepackage{bm}
\usepackage{color}
\journal{Physics Letters B}

\begin{document}
\begin{frontmatter}

\title{Decoding the nuclear symmetry energy event-by-event in heavy-ion collisions with machine learning}
\author[1]{Yongjia Wang}
\author[1,2]{Zepeng Gao}
\author[3]{Hongliang L\"{u}}
\author[1,4]{Qingfeng Li\corref{corr1}}\ead{liqf@zjhu.edu.cn}

\cortext[corr1]{Corresponding author}
\address[1]{School of Science, Huzhou University, Huzhou 313000, China}
\address[2]{Sino-French Institute of Nuclear Engineering and Technology, Sun Yat-sen University, Zhuhai 519082, China}
\address[3]{HiSilicon Research Department, Huawei Technologies Co., Ltd., Shenzhen 518000, China}
\address[4]{Institute of Modern Physics, Chinese Academy of Sciences, Lanzhou 730000, China}

\begin{abstract}
Inferences of the nuclear symmetry energy from heavy-ion collisions are currently based on the comparison of measured observables and transport model simulations. Only the expectation values of observables over all considered events are used in these approaches, however, observables can be obtained event-by-event both in experiments and transport model simulations.
By using the light gradient boosting machine (LightGBM), a modern machine-learning algorithm, we present a framework for inferring the density-dependent nuclear symmetry energy from observables in heavy-ion collisions on the event-by-event analysis. The ultrarelativistic quantum molecular dynamics (UrQMD) model simulations are used as training data. The symmetry energy slope parameter extracted with LightGBM event-by-event from test data also by UrQMD has an average spread of approximately 30~MeV from the truth, and is found to be robust against variations in model parameters. In addition, LightGBM can identify features that have the greatest effect on the physics of interest, thereby offering valuable insights. Our study suggests that the present framework can be a powerful tool and may offer a new paradigm to study the underlying physics in heavy-ion collisions.

\end{abstract}


\end{frontmatter}

$Introduction$.-Nuclear symmetry energy, which quantifies the difference in binding energy per nucleon between isospin symmetric and asymmetric nuclear matter, is of crucial importance to the structure of nuclei (e.g., the neutron skin thicknesses and dipole polarizabilities), the dynamics of heavy-ion collision with radioactive beams (e.g., yield ratios between isospin partners), as well as properties of compact stars (e.g., neutron star tidal deformability and its mass-radius relationship) \cite{Li:2008gp,Tsang:2012se,meng,Oertel:2016bki,Li:2019xxz,Li:2021thg,Reinhard:2021utv,Huth:2021bsp}. Deducing how the nuclear symmetry energy varies with density $E_{\rm sym}(\rho)$ from theories and experiments is one of the hot topics in nuclear physics and nuclear astrophysics. Heavy-ion collisions (HICs) with various species and energies offer a unique opportunity to access the nuclear symmetry energy at densities away from the saturation density ($\rho_0$). However, $E_{\rm sym}(\rho)$ cannot be measured directly in experiment; usually it is deduced from the comparison between transport model simulations and experimental measurements \cite{Ono:2019jxm,Xu:2019hqg,Colonna:2020euy,ma}. Consequently, constraints on $E_{\rm sym}(\rho)$ from different models and experimental data diverge, especially above $\rho_0$. There are a few reasons for this: (I) different assumptions and coding techniques are used in transport models--to disentangle these issues, the Transport Model Evaluation Project (TMEP) has been pursued since several years ago \cite{SpRIT:2020blg,Xu:2016lue,Zhang:2017esm,Ono:2019ndq,Colonna:2021xuh,hermann}; (II) different observables which may reflect properties of $E_{\rm sym}(\rho)$ at different densities are used to constrain $E_{\rm sym}(\rho)$ \cite{Russotto:2016ucm,Wang:2020dru,Liu:2020jbg,SRIT:2021gcy};
(III) the nuclear symmetry energy effects on observables in HICs can be easily concealed by large fluctuations caused by stochastic nucleon-nucleon collisions and by initial state fluctuations \cite{Zhang:2004kq,Wang:2021sdu,Li:2022mni}.

To identify fingerprints of $E_{\rm sym}(\rho)$ on observables in HICs is one way to probe $E_{\rm sym}(\rho)$. Traditionally, averaged values of observables over all considered events are usually used to constrain $E_{\rm sym}(\rho)$ from the comparison of experimental measurements and transport model simulations. In this context, the extent of data used in traditional method is limited. Indeed, observables can be obtained event-by-event both in experiments and transport model simulations. Event-by-event data in HICs may consist of a huge amount of data of different observables for each event. These event-by-event data carries a lot of information about relevant physics and correlations among observables. Analyzing these event-by-event data and mining relevant information is a big challenge in traditional analysis. Can we build a framework that uses event-by-event data to infer $E_{\rm sym}(\rho)$ with modern data science techniques? Moreover, in traditional analysis, unsettled parameters (assumptions) of transport models have been found to affect the averaged values resulting in the divergence of the constrained $E_{\rm sym}(\rho)$ \cite{Coupland:2011px,Zhang:2015xna,Cozma:2017bre,Morfouace:2019jky,Wang:2020dru,SpRIT:2020blg,Cozma:2021tfu,SpRIT:2021dvt}. Therefore, it is desirable to infer $E_{\rm sym}(\rho)$ in a model-independent way.

Machine learning has been proven powerful for analyzing complex data in many branches of science \cite{Pang:2016vdc,Bedaque:2021bja,Mehta:2018dln,Carleo:2019ptp,Boehnlein:2021eym,book,Du:2021pqa}. In this work, by using a modern machine-learning algorithm, we present a machine learning framework for inferring $E_{\rm sym}(\rho)$ with event-by-event data, and the framework is found to be robust against variations in model parameters. This may open a new venue to study the underlying physics in HICs.


\begin{figure}[t]
\begin{center}
\includegraphics*[scale=0.30]{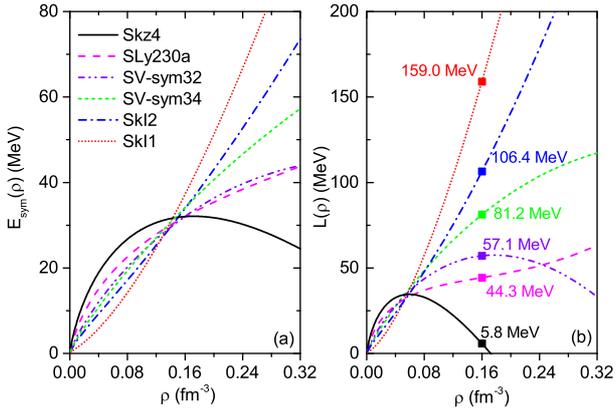}
\caption{(Color online) The nuclear symmetry energy and its slope parameter are plotted as a function of density. Numbers listed in the right panel are the slope parameters at the saturation density.  }
\label{fig1}
\end{center}
\end{figure}

$Methodology$.-Supervised learning is one of the machine learning techniques to map the relationship between input features and the output. Identifying $E_{\rm sym}(\rho)$ from observables in HICs can be designed as a supervised learning task. One way to generate labeled data (observables that have been tagged with different $E_{\rm sym}(\rho)$) is theoretical simulation. For this purpose, we use the ultrarelativistic quantum molecular dynamics (UrQMD) model. UrQMD has been widely employed for investigating HICs over a broad energy range \cite{SAB,BLE,Bleicher:2022kcu,qfli1}. In presently used UrQMD model, the symmetry potential is derived from the Skyrme potential energy density functional \cite{Zhang:2006vb,Wang:2013wca,FOP-wyj,FOP-zyx}, while the isoscalar component of the mean field potential inherits from the widely used soft and momentum dependent potential (SM, with the incompressibility $K_0$=200 MeV) in QMD-like models \cite{Aichelin:1991xy,Hartnack:1997ez}. Six different Skyrme interactions (Skz4, SLy230a, SV-sym32, SV-sym34, SkI2, and SkI1 \cite{Dutra:2012mb}) which yield very different $E_{\rm sym}(\rho)$ are considered in the present work. $E_{\rm sym}(\rho)$ and its slope parameter $L(\rho)=3\rho \left (\partial{E_{\rm sym}(\rho)}/\partial{\rho} \right )$ as a function of density are plotted in Fig.~\ref{fig1} (a) and (b), respectively. It can be seen that $L$($\rho$) spans a wide range of values, fully covering the present uncertainty on $E_{\rm sym}(\rho)$.

The SM nuclear potential, with the FP4 parametrization \cite{Wang:2013wca} for the in-medium nucleon-nucleon cross section and an isospin-dependent minimum spanning tree (isoMST) method for cluster recognition, is applied as the default parameter set. In isoMST, two protons (two neutrons or neutron-proton pair) are considered to belong to the same fragment if their relative distance and momentum are smaller than $R_{0}^{pp}$ ($R_{0}^{nn}$ or $R_{0}^{np}$) and $P_{0}$, respectively. The parameters adopted in the default parameter set are $R_{0}^{pp}$=2.8 fm, $R_{0}^{nn}$=$R_{0}^{np}$=3.8 fm, and $P_{0}$=0.25 GeV/c. In the present work, the training data consist of data generated with UrQMD using the default parameter set. $Testdata1$ is consistent with the training data.
To validate the generalizability of trained machine learning algorithm, other parameter sets of UrQMD are also used to generate testing data, i.e., the free nucleon-nucleon cross section ($Testdata2$), larger parameters of $R_{0}^{pp}$=3.5 fm, $R_{0}^{nn}$=$R_{0}^{np}$=4.5 fm and $P_{0}$=0.3 GeV$/$c in isoMST  ($Testdata3$), a hard and momentum dependent potential (HM,  with the incompressibility $K_0$=380 MeV, $Testdata4$). We note here that, when varying model parameters in UrQMD, the magnitudes of observables are visibly influenced. For example, the yield of free protons in the case of $Testdata1$ is about 10\% larger than that in $Testdata3$. The elliptic flow obtained with HM is about 50\% stronger than that obtained with SM. When the nuclear symmetry energy is varied, the variations in observables are usually small. In practice, Au+Au collisions at beam energy of 0.4 GeV$/$nucleon with impact parameter $b$=5 fm are simulated; the number of events for generating training data and each testing data are 360000 and 50000 for each $E_{\rm sym}(\rho)$, respectively. We have checked that the results are not changed with further increasing the number of events.

\begin{figure}[t]
\begin{center}
\includegraphics*[scale=0.4]{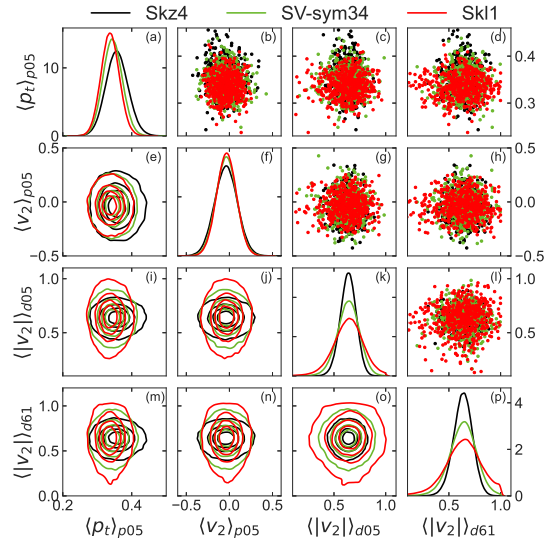}
\caption{(Color online) The diagonal plots show the distribution of each observable obtained with Skz4, SV-sym34, and SkI1. Upper off-diagonal scatter plots  display the correlation between the horizontal and vertical labeled observables. Each scatter point denotes one single event; total 1000 events are displayed. Lower off-diagonal plots are the corresponding contour plots.}
\label{scat}
\end{center}
\end{figure}

For data generation, 30 features (observables) related to the momenta ($p_x$, $p_y$, and $p_z$) of produced free protons and deuterons are used. They are $\langle p_t \rangle$, $\langle |p_x| \rangle$, $\langle |p_y| \rangle$, $\langle v_2 \rangle$, $\langle |v_1| \rangle$, $\langle |v_2| \rangle$, and $\langle |v_3| \rangle$ of free protons and deuterons within rapidity windows $|y_0| < 0.5$ and $0.6 < |y_0| < 1.0$, and additionally $\langle |p_z| \rangle$ of free deuterons. Here, $p_{t}=\sqrt{p_{x}^{2}+p_{y}^{2}}$, $v_1$=$p_x / p_t$, $v_2$=$(p_{x}^{2}-p_{y}^{2})/p_{t}^{2}$, and $v_3$=$(4p_{x}^{3}-3p_xp_{t}^{2})/p_{t}^{3}$. The angle brackets indicate an average over all considered particles in each event. The normalized rapidity is defined as $y_{0}=y_{z}/y_{pro}$ with $y_{pro}$ being the projectile rapidity in the center-of-mass frame. Observables with sub-index $p05$ ($p61$) and $d05$ ($d61$) referring to the results for free protons and deuterons within rapidity window $|y_0| < 0.5$ ($0.6 < |y_0| < 1.0$), respectively. To reveal how nuclear symmetry potential influence these observables, the distribution, scatter plot, and contour plot of $\langle p_t \rangle$$_{p05}$, $\langle v_2 \rangle$$_{p05}$, $\left\langle \left|v_2\right|\right\rangle$$_{d05}$, and $\left\langle \left|v_2\right|\right\rangle$$_{d61}$ are displayed as example in Fig. \ref{scat}. From the univariate distribution of each
observable, as showed in the diagonal of Fig. \ref{scat}, one can infer the effects of nuclear symmetry energy on the mean value (over all considered events). This is conventional analysis that has been widely used to deduce the underlying physics from HICs. For example, the mean $p_t$ of free protons within $|y_0|\le0.5$ obtained with Skz4 is about 0.357 GeV$/$c while that obtained with SkI1 is about 0.338 GeV$/$c. SkI1 leads to a smaller $p_t$ for protons because of a stronger attraction from a stiffer $E_{\rm sym}(\rho)$.
One may constrain $E_{\rm sym}(\rho)$ by comparing experimentally measured $\langle p_t \rangle$ to theoretical calculations with different $E_{\rm sym}(\rho)$. However, the mean $p_t$ of free protons are apparently also affected by other model parameters \cite{Russotto:2011hq}, so such a constrained $E_{\rm sym}(\rho)$ may be strongly biased.
In the scatter and contour plots of Fig.~\ref{scat}, one sees the effects of $E_{\rm sym}(\rho)$ on a two-dimensional plane. The results obtained from different $E_{\rm sym}(\rho)$ largely overlap because the effect of symmetry potential is relatively weak. Usually, besides using one observable, two correlated observables (e.g., the $p_t$ dependent $v_2$ ratio of isospin partners) are used to probe $E_{\rm sym}(\rho)$ \cite{Russotto:2011hq,Russotto:2016ucm}. In traditional analysis, it is almost impossible to use simultaneously more than three observables. Machine learning has been proved useful for extracting information or patterns from complex data, thus one may expect machine learning algorithms to be able to discriminate $E_{\rm sym}(\rho)$ from mutli-observables.

To do so, the light gradient boosting machine (LightGBM),  a decision-tree-based algorithm, is employed. LightGBM has higher accuracy, faster training efficiency, as well as stronger ability to handle large-scale data \cite{LightGBM}, thus it has been widely used and has achieved state-of-the-art performances in many tasks. Moreover, comparing to other machine learning algorithms, LightGBM yields a higher interpretability because of its decision-tree nature. In our previous works, the strong ability of LightGBM to refine nuclear mass models and determine impact parameter of HICs has been demonstrated \cite{Gao:2021eva,Li:2020qqn,Li:2021plq}. All parameters in LightGBM are set to their default values; we have checked that the results are insensitive to parameters in LightGBM. To combat overfitting, the so-called $L1$ and $L2$ regularization \cite{LightGBM} are implemented. In addition, testing datasets which constitute unseen data and data generated with different parameter sets in UrQMD are used to further prevent overfitting.

\begin{figure}[t]
\begin{center}
\includegraphics*[angle=0,width=0.4\textwidth]{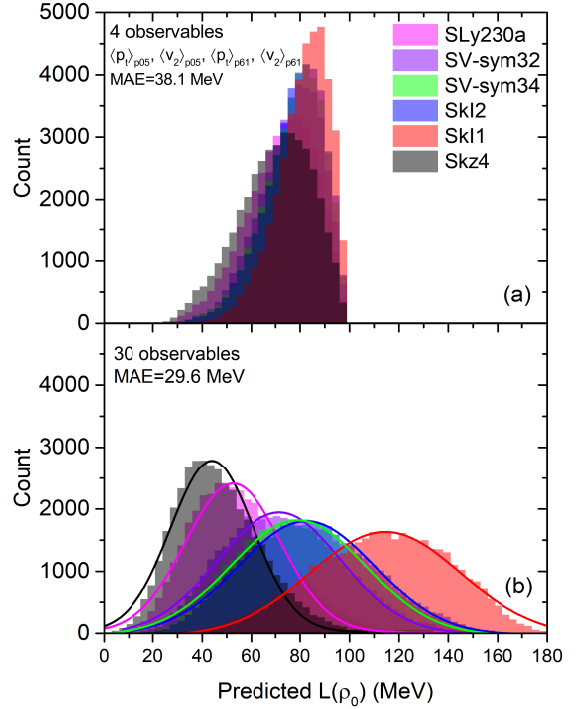}
\caption{(Color online) The distributions of the predicted slope parameter $L(\rho_0)$ from $TestData1$. Upper and lower panels are the results obtained with 4 and 30 observables, respectively.
Curves are Gaussian fits to the distributions.}
\label{chi2}
\end{center}
\end{figure}

\begin{table*}[htbp]
\caption{\label{tab:table1} The mean values of predicted $L(\rho_0)$ and their standard deviation $\sigma$ obtained with Gaussian fit. All units are in MeV.}
\setlength{\tabcolsep}{1.2pt}
\begin{tabular}{lc |cc |cc |cc |cc}
&  & \multicolumn{2}{c|}{$Testdata1$ (MAE=29.6)}  &\multicolumn{2}{c|}{$Testdata2$ (MAE=29.4)} &\multicolumn{2}{c|}{$Testdata3$ (MAE=29.4)} &\multicolumn{2}{c}{$Testdata4$ (MAE=27.8)}\\
\hline
 & $L^{\rm true}(\rho_0)$ & $\langle L^{\rm pred}(\rho_0) \rangle$  & $\sigma$  & $\langle L^{\rm pred}(\rho_0) \rangle$   & $\sigma$  & $\langle L^{\rm pred}(\rho_0) \rangle$ & $\sigma$  & $\langle L^{\rm pred}(\rho_0) \rangle$  & $\sigma$    \\
Skz4     & 5.8              & 44.1           & 16.8   & 43.3  & 16.1    & 38.4    & 16.4    & 48.0   & 17.0          \\
SLy230a  & 44.3             & 52.3           & 19.4   & 51.3  & 17.5    & 47.3    & 19.0    & 58.7   & 20.2          \\
SV-sym32 & 57.0             & 71.3           & 25.1   & 69.1  & 23.2    & 66.6    & 25.3    & 82.9   & 25.8          \\
SV-sym34 & 81.2             & 78.8           & 27.2   & 76.6  & 24.8    & 73.9    & 27.2    & 93.0   & 27.6          \\
Skl2     & 106.4            & 82.8           & 27.9   & 79.6  & 25.7    & 77.7    & 28.1    & 98.6   & 28.2          \\
Skl1     & 159.0            & 114.9          & 29.7   & 110.3 & 29.8    & 109.7   & 31.5    & 140.8  & 22.6          \\
\hline
\end{tabular}\label{reg}
\end{table*}

$Results$.-Both the training and testing data (i.e., $Testdata1$) generated with the same parameter set in UrQMD are tested, and the distribution of the predicted $L(\rho_0)$ is plotted in Fig. \ref{chi2}. First, similar to traditional analysis,
if only the $p_t$ dependent $v_2$ observables are used, such as those listed in Fig.~\ref{chi2} (a), all of the predicted $L(\rho_0)$ approach to the global-averaged value [(5.8+44.3+57.0+81.2+106.4+159.0)/6=75.6 MeV] and the mean absolute error (MAE), which is the absolute difference between the true and the predicted $L(\rho_0)$, is as large as 38.1 MeV. Second, with 30 observables [see Fig. \ref{chi2} (b)], the distributions of predicted $L(\rho_0)$ are fairly separated and the MAE drops to 29.6 MeV.
We note that this MAE value is larger than our previous result~\cite{Wang:2021xbb}. This is because the present study is event-by-event whereas our previous study treated the average over 100 events as one sample.
We have checked, by using averaged observables over 100 events, that the MAE drops to 13.4 MeV with the present method. It is conceivable, with more variables included, that the widths of the distributions in Fig.~\ref{chi2} (b) would narrow further and the peaks would be more cleanly separated.
This demonstrates clearly the power of machine learning which can incorporate practically unlimited number of variables, which would be completely out of the question for a traditional analysis.

The mean value of the predicted $L(\rho_0)$, its standard deviation, and the MAE for $Testdata1$ using 30 variables are listed in Table~\ref{reg} for the nuclear potentials we have studied.
It is noteworthy for the cases of SLy230a, SV-sym32, SV-sym34, and SkI2 that the difference between the mean values of predicted $L(\rho_0)$ and the true values used in corresponding events are smaller than 24 MeV, while the differences in the cases of Skz4 and SkI1 (very soft and stiff ones) are as large as about 40 MeV. It is known that test dataset on extremes of the spectrum of the training datasets would yield event-by-event results significantly deviating from the truth. With a given real dataset, one would reiterate the procedure by including a wide spectrum of training data well embracing the test results.
In reality, in the case of symmetry energy we study, it is known that $E_{\rm sym}(\rho)$ is unlikely to be very soft or stiff \cite{Li:2021thg,Essick:2021kjb,SRIT:2021gcy,Yue:2021yfx,meng2}, thus the iteration would quickly converge.

To illustrate the generalizability of the trained LightGBM model, we tabulate in Table~\ref{reg} results for all testing datasets. Values of MAE from $Testdata2$, $Testdata3$, and $Testdata4$ are similar to that from $Testdata1$. This implies  that the trained LightGBM model is reliable even for inputs generated by UrQMD with different parameter sets, indicating small model-dependent systematic uncertainties. This suggests that the use of LightGBM to identify $E_{\rm sym}(\rho)$ from more than a handful observables has considerable potential.

Experimental data are often a mixture in impact parameter ($b$) which cannot be precisely determined. We thus test the trained LightGBM using data generated by UrQMD with two ranges of $b$=0-2 fm and $b$=2-4 fm. The distributions of the predicted $L(\rho_0)$ are similar to those  in Fig.\ref{chi2}, with the corresponding MAE of 32.7 MeV and 31.2 MeV. These values are only slightly larger than that for fixed $b$ dataset, suggesting a strong potential of LightGBM in extracting $E_{\rm sym}(\rho)$ from real experimental data. This is non-trivial as varying impact parameter can sometimes result in dramatic changes in observables.


\begin{figure*}[htbp]
\begin{center}
\includegraphics[angle=0,width=0.8\textwidth]{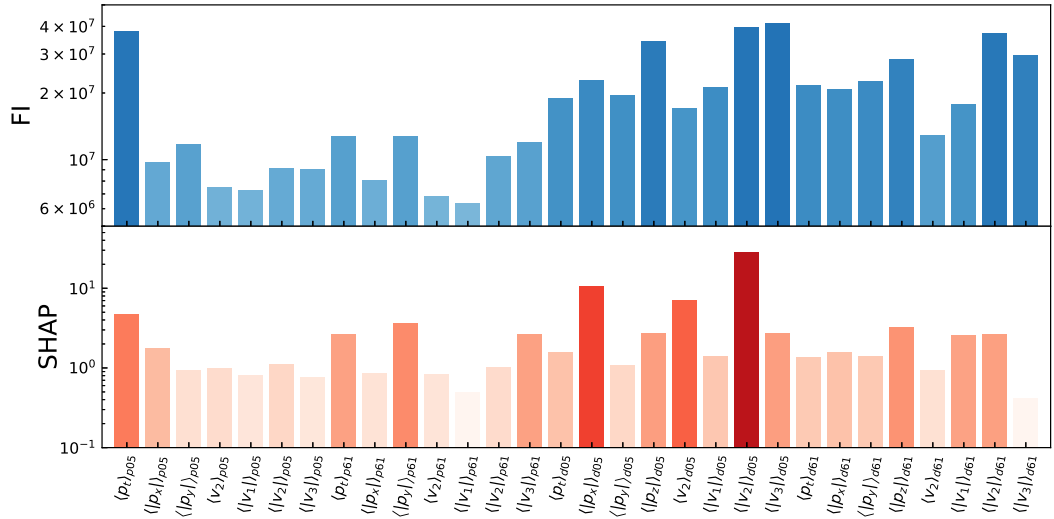}
\caption{(Color online) The FI (upper) and SHAP (lower) values for each feature (variable) used in our study.
}
\label{rel}
\end{center}
\end{figure*}

Comparing to other deep learning algorithms, often treated as block boxes, LightGBM has an excellent degree of explainability because of its decision-tree-based nature. This is important as explainability often improves our
knowledge about the relationship between the input features and the output. The $Feature$$\_$$importance$ (FI) technology
of LightGBM and Shapley additive explanations (SHAP) are two popular feature attribution methods to identify the most important
features that drive predictions \cite{LightGBM,SHAP}. The FI and SHAP values for each feature obtained with the above methods are displayed in Fig.~\ref{rel}.
These two methods utilize different techniques to characterize the importance~\cite{LightGBM,SHAP}; although not equal, the identified important features are well correlated between the two methods. In our particular study, deuteron related features exhibit higher importance than protons in the determination of $L(\rho_0)$. This finding suggests that one can concentrate on deuteron-related observables to probe $E_{\rm sym}(\rho)$.
It can be understood from the fact that the distribution width of each deuteron-related observable is sensitive to $E_{\rm sym}(\rho)$, as displayed in Fig. \ref{scat}. LightGBM is able to use this information to identify $E_{\rm sym}(\rho)$ on an event-by-event basis, which cannot be done in conventional analysis. It demonstrates the strong ability of LightGBM to decode information from complex data.

$Conclusion$.-We have shown that the fingerprints of the nuclear symmetry energy can be decoded from a large set of  observables in heavy-ion collisions on an event-by-event basis by the trained machine learning algorithm LightGBM.  Trained with data generated by the UrQMD model, LightGBM is able to predict the symmetry energy slope parameter $L(\rho_0)$ with a mean absolute error of approximately 30~MeV. The results are robust against UrQMD model parameters in the nuclear potential, in-medium cross section, and cluster recognition.
It is found that LightGBM works well with mixed impact parameters which ensures its applicability to real data analysis.
The present framework can be generalized to studies of other physics in heavy-ion collisions, for example the extraction of
the nuclear equation of state, the in-medium correction to the nucleon-nucleon cross section, the quark gluon plasma shear viscosity.
It may open a new paradigm for studying the underlying physics in heavy-ion collisions.

We thank Fuqiang Wang for a careful reading of the
manuscript and valuable communications. We acknowledge fruitful discussions with the TMEP group, M. B. Tsang, Yingxun Zhang, Kai Zhou, Jan Steinheimer, Yilun Du, Nan Su, Longgang Pang, Jun Su, Long Zhu, and Horst Stoecker.
The authors are grateful to the C3S2 computing center in Huzhou University for calculation support. The work is supported in part by the National Natural Science Foundation of China (Nos. U2032145, 11875125, and 12147219), the National Key Research and Development Program of China under Grant No. 2020YFE0202002. The contributions of Hong-Liang L\"{u} are non-Huawei achievements

\end{document}